\documentclass[journal=jacsat,manuscript=article]{achemso}
\usepackage{graphicx}
\usepackage{dcolumn}
\usepackage{bm}
\usepackage{hyperref}
\usepackage{color}
\usepackage{xcolor}
\usepackage{ulem}
\usepackage{amsmath}
\usepackage[english]{babel}
\hypersetup{
    colorlinks=true,
		citecolor=blue,
    linkcolor=blue,
    urlcolor=blue,
}


\DeclareGraphicsExtensions{.pdf,.png,.jpg}
\author{Rui Lou}
\email{lourui@lzu.edu.cn}
\affiliation{School of Physical Science and Technology, Lanzhou University, Lanzhou 730000, China}
\alsoaffiliation{Leibniz Institute for Solid State and Materials Research, IFW Dresden, 01069 Dresden, Germany}
\alsoaffiliation{Helmholtz-Zentrum Berlin f{\"u}r Materialien und Energie, Albert-Einstein-Stra{\ss}e 15, 12489 Berlin, Germany}
\alsoaffiliation{Joint Laboratory ``Functional Quantum Materials" at BESSY II, 12489 Berlin, Germany}

\author{Oleksandr Suvorov}
\affiliation{Leibniz Institute for Solid State and Materials Research, IFW Dresden, 01069 Dresden, Germany}
\alsoaffiliation{\textcolor{black}{G. V. Kurdyumov Institute for Metal Physics of the N.A.S. of Ukraine, 03142 Kyiv, Ukraine}}

\author{Hans-Joachim Grafe}
\affiliation{Leibniz Institute for Solid State and Materials Research, IFW Dresden, 01069 Dresden, Germany}

\author{Andrii Kuibarov}
\affiliation{Leibniz Institute for Solid State and Materials Research, IFW Dresden, 01069 Dresden, Germany}

\author{Maxim Krivenkov}
\affiliation{Helmholtz-Zentrum Berlin f{\"u}r Materialien und Energie, Albert-Einstein-Stra{\ss}e 15, 12489 Berlin, Germany}

\author{Oliver Rader}
\affiliation{Helmholtz-Zentrum Berlin f{\"u}r Materialien und Energie, Albert-Einstein-Stra{\ss}e 15, 12489 Berlin, Germany}

\author{Bernd B{\"u}chner}
\affiliation{Leibniz Institute for Solid State and Materials Research, IFW Dresden, 01069 Dresden, Germany}
\alsoaffiliation{Institute for Solid State and Materials Physics, TU Dresden, 01062 Dresden, Germany}

\author{Sergey Borisenko}
\email{s.borisenko@ifw-dresden.de}
\affiliation{Leibniz Institute for Solid State and Materials Research, IFW Dresden, 01069 Dresden, Germany}

\author{Alexander Fedorov}
\email{a.fedorov@ifw-dresden.de}
\affiliation{Leibniz Institute for Solid State and Materials Research, IFW Dresden, 01069 Dresden, Germany}
\alsoaffiliation{Helmholtz-Zentrum Berlin f{\"u}r Materialien und Energie, Albert-Einstein-Stra{\ss}e 15, 12489 Berlin, Germany}
\alsoaffiliation{Joint Laboratory ``Functional Quantum Materials" at BESSY II, 12489 Berlin, Germany}

\title[An \textsf{achemso} demo]
  {Suppression of nematicity by tensile strain in multilayer FeSe/SrTiO$_3$ films}
\begin{document}	
	\begin{abstract}
      \textcolor{black}{The nematicity in multilayer FeSe/SrTiO$_3$ films has been previously suggested to be enhanced with decreasing film thickness. Motivated by this, there have been many discussions about the competing relation between nematicity and superconductivity. However, the criterion for determining the nematicity strength in FeSe remains highly debated. The understanding of nematicity and its relation to superconductivity in FeSe films is therefore still controversial.}
      Here, we fabricate \textcolor{black}{multilayer FeSe/SrTiO$_3$ films} using molecular beam epitaxy and study the nematic properties by combining angle-resolved photoemission spectroscopy, nuclear magnetic resonance, and scanning tunneling microscopy experiments.
      \textcolor{black}{We unambiguously demonstrate that, near the interface, the nematicity is suppressed by the SrTiO$_3$-induced tensile strain;
      in the bulk region further away from the interface, the strength of nematicity recovers to the bulk value.} Our results not only solve the controversy about the nematicity in multilayer FeSe films, but also offer valuable insights into the relationship between nematicity and superconductivity.
\end{abstract}

\maketitle

\subsection{Introduction}
Electronic nematicity, which breaks rotational symmetry but preserves translational symmetry of lattice \cite{KivelsonSA1998}, has been widely observed in cuprates \cite{AndoY2002,DaouR2010,LawlerMJ2010,WuJ2017} and iron-based superconductors \cite{ChuangTM2010,ChuJH2010,ChuJH2012,KuoHH2016,YiM2011,ZhangYNaFeAs,FernandesRM2014NP}.
In iron-based superconductors, nematicity is usually accompanied by a tetragonal-to-orthorhombic structural transition \cite{JohnstonDC2010}. Besides,
\textcolor{black}{the collinear antiferromagnetism} sets in at or just below the nematic transition in iron pnictides \cite{CruzCdl2008,ZhaoJ2008},
\textcolor{black}{the complicated coupling between lattice, spin, and orbital degrees of freedom \cite{KasaharaS2012,LuX2014,YiM2011} renders the origin of nematicity controversial \cite{FangC2008,XuCK2008,LeeCC2009,ChenCC2010}.} FeSe is an attractive system \textcolor{black}{to explore the effect of nematicity disentangled from that of magnetic order,} as it exhibits nematicity below $T_{\rm nem}$ $\sim$ 90 K without long-range magnetic order at any temperature \cite{McQueenTM2009,MedvedevS2009,BaekSH2015,NakayamaK2014,WatsonMD2015,BohmerAE2015}.
FeSe also provides a versatile platform for studying the interplay of nematicity and superconductivity since \textcolor{black}{the superconductivity therein
is highly tunable
\cite{GuoJG2010,LuXF2015,DongXL2015PRB,DongXL2015,ShahiP2018,WangQY2012,TanS2013,HeSL2013,GeJF2015}.}

The relationship between nematicity and superconductivity is under hot debate in \textcolor{black}{FeSe crystals.}
Their competition is proposed in an electron irradiation experiment, where the disorder enhances $T_c$ while
lowers $T_{\rm nem}$ \cite{TeknowijoyoS2016}, as hydrostatic pressure \cite{WangPS2016} and uniaxial strain do \cite{GhiniM2021,BartlettJM2021}. \textcolor{black}{Whereas, their cooperative effect is suggested by the sizeable enhancement of $T_c$ with an enlarged orthorhombic distortion below $T_c$
upon increasing sulfur doping in FeSe \cite{WangLR2016}.} \textcolor{black}{Contrary to a direct correlation,}
thermal-expansion measurements find a lack of coupling between nematicity and superconductivity \cite{BohmerAE2013PRB}.

\textcolor{black}{Thus the origin of nematicity in FeSe remains highly interesting, but its effect on the electronic structure is controversial. \textcolor{black}{In FeSe crystals, due to the complicated band structure and multiorbital character around Brillouin zone (BZ) corner, the energy scale of the zone-corner $d_{xz}$/$d_{yz}$ splitting has been hotly debated. It was reported to be either $\sim$50 meV \cite{NakayamaK2014,WatsonMD2015,ZhangP2015,ShimojimaT2014,YiM2019PRX} or degenerate \cite{WatsonMD2016PRB,WatsonMD2017NJP},}} questioning whether the bands at BZ corner are appropriate for understanding the nematicity.
\textcolor{black}{Instead, the relatively simple band structure around BZ center provides a remarkable platform, by which the nematic energy scale of $\sim$15-20 meV agrees with the $T_{\rm nem}$ scale \cite{WatsonMD2016PRB,FedorovA2016,WatsonMD2017NJP,WatsonMD2015}.
\textcolor{black}{In multilayer FeSe/SrTiO$_3$ films, it has been previously suggested that the strength of nematicity in thinner FeSe films is stronger than that in FeSe crystals, and it further enhances with decreasing thickness \cite{TanS2013,Peng2015CPB,WenC2016NC,20ML2017PRB,XJZhou2016CPL}. However, in these studies, the nematicity strength was determined by the band splitting around BZ corner, the validity of this criterion has been questioned in FeSe crystals \cite{ColdeaA2018review}. Therefore, whether the nematicity is enhanced in thinner FeSe films is still an open question.}}


\textcolor{black}{Herein} we present combined angle-resolved photoemission spectroscopy (ARPES), $^{77}$Se nuclear magnetic resonance (NMR), and scanning tunneling microscopy (STM) studies on multilayer FeSe/SrTiO$_3$ films.
\textcolor{black}{NMR measurements show the coexistence of single-domain and twinned-domain structures in 210-monolayer (ML) film; as 15-ML film forms a single domain revealed by STM, we suggest that the former structure is near the interface and the latter one is from the bulk region further away from the interface under noticeable and negligible tensile strain, respectively. Consistently, in 15-ML film, the single circular hole Fermi surface (FS) and the temperature-independent $d_{xz}$/$d_{yz}$ splitting around $\bar{\Gamma}$ indicate the absence of nematicity; in 115-ML film, the splitting of outer hole band and the temperature evolutions of $d_{xz}$ and $d_{yz}$ orbitals at $\bar{\Gamma}$ validate the emergence of bulk-like nematicity.}

\begin{figure*}[h!t]
  \begin{center}
  \includegraphics[width=10.5cm]{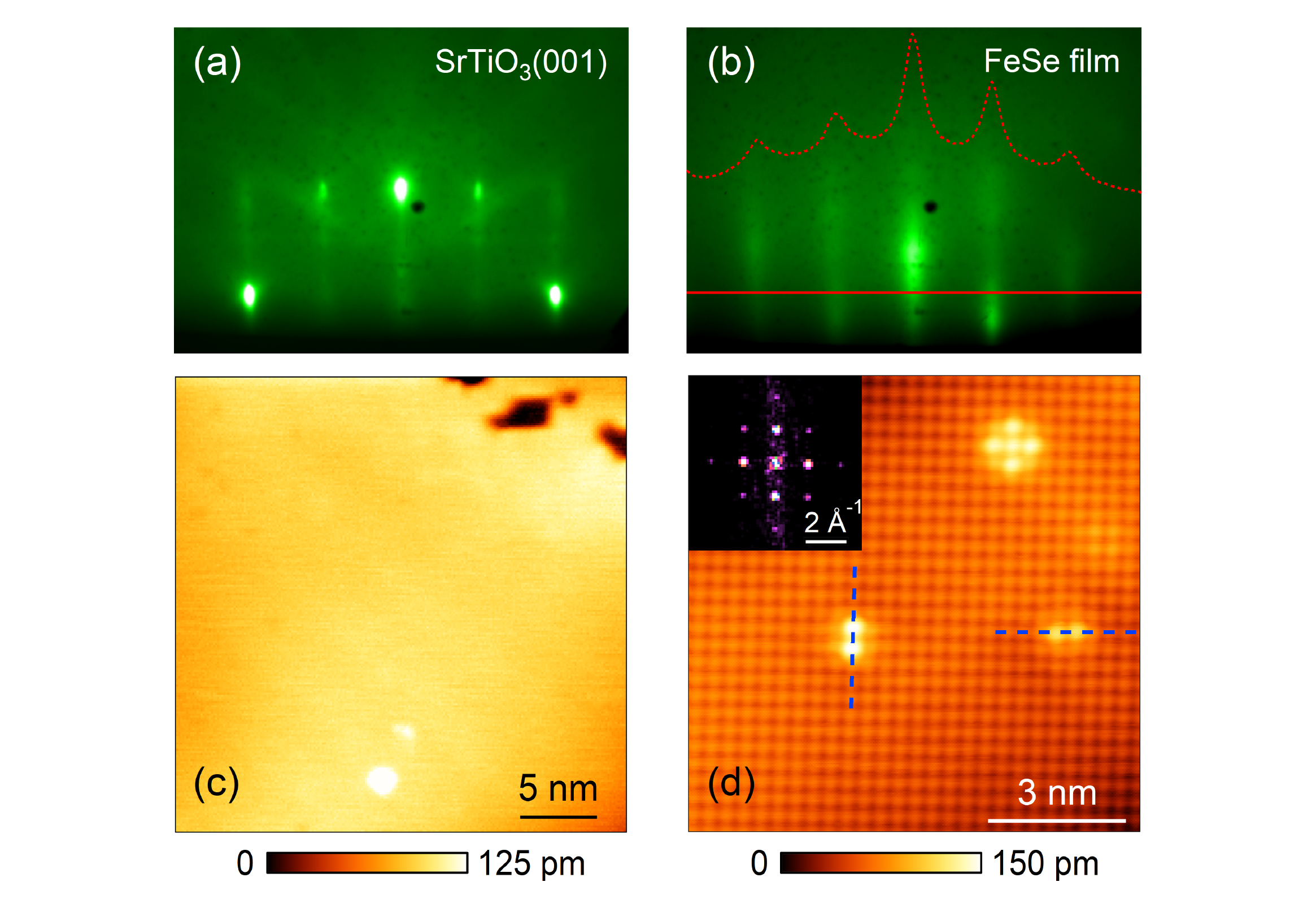}
  \end{center}
  \caption{\textbf{Surface and structural characterization.}
  (a),(b) RHEED images of SrTiO$_3$ substrate and 15-ML film with incident beam along the \textcolor{black}{[110]} directions, respectively.
  \textcolor{black}{The RHEED intensity curve (red dashed curve) is taken along the red solid line. The single set of diffraction peaks reflect the high quality of FeSe film.}
  (c) STM topographic image of 15-ML film (30 $\times$ 30 nm$^2$, $V_s$ = 50 mV, $I$ = 50 pA).
  (d) Atomically resolved STM topography of 15-ML film (10 $\times$ 10 nm$^2$, $V_s$ = 50 mV, $I$ = 100 pA).
  \textcolor{black}{Blue dashed lines indicate the orientations of Fe vacancies.}
  Inset: Fourier transform of the defect-free regions in (d).
  }
\end{figure*}

\subsection{Results}
Multilayer FeSe films were fabricated on either Nb:SrTiO$_3$(001) (for ARPES and STM) or SrTiO$_3$(001) (for NMR) substrates.
The surface and structural characterization of 15-ML film are presented in Fig. 1.
\textcolor{black}{Figures 1(a) and 1(b) show the reflection high-energy electron diffraction (RHEED) patterns of SrTiO$_3$ substrate and 15-ML film at room temperature, respectively.}
A typical STM topographic image of 15-ML film at $\sim$4.5 K is displayed in Fig. 1(c), where the surface consists of a single crystallographic domain. The  single-domain feature is observed in a number of scan areas over the sample surface, \textcolor{black}{unlike the presence of twinned domains with twin boundaries  previously reported in thicker FeSe films \cite{LiW2017NP} and FeSe crystals \cite{WatashigeT2015}.} The atomic-resolution STM image and its Fourier transform in Fig. 1(d) reveal a square-like lattice of the topmost Se atoms.
As in previous studies \cite{LiW2017NP,WatashigeT2015,SongScience2011,SongPRB2011}, the Fe (dumbbell-like) and Se (cross-like) vacancies are observed.
\textcolor{black}{Recently, the appearance of stripe patterns pinned around the Fe vacancies has been proposed as a signature of the nematicity-related orthorhombicity in FeSe films \cite{LiW2017NP}; the interaction between the stripes and Fe vacancies has been suggested to induce strong distortions of the latter, that the directions of the dumbbell shapes show obvious deviation from that of the Se-Se lattice \cite{LiW2017NP}. However, in 15-ML film [Fig. 1(d)],
we do not observe such stripe patterns near the dumbbell-like impurities; as indicated by the blue dashed lines, the Fe vacancies are well aligned along the directions of the Se-Se lattice, showing no evident distortions. These facts indicate the absence of orthorhombicity in 15-ML film.}

\begin{figure*}[h!t]
	\begin{center}
  \includegraphics[width=10.5cm]{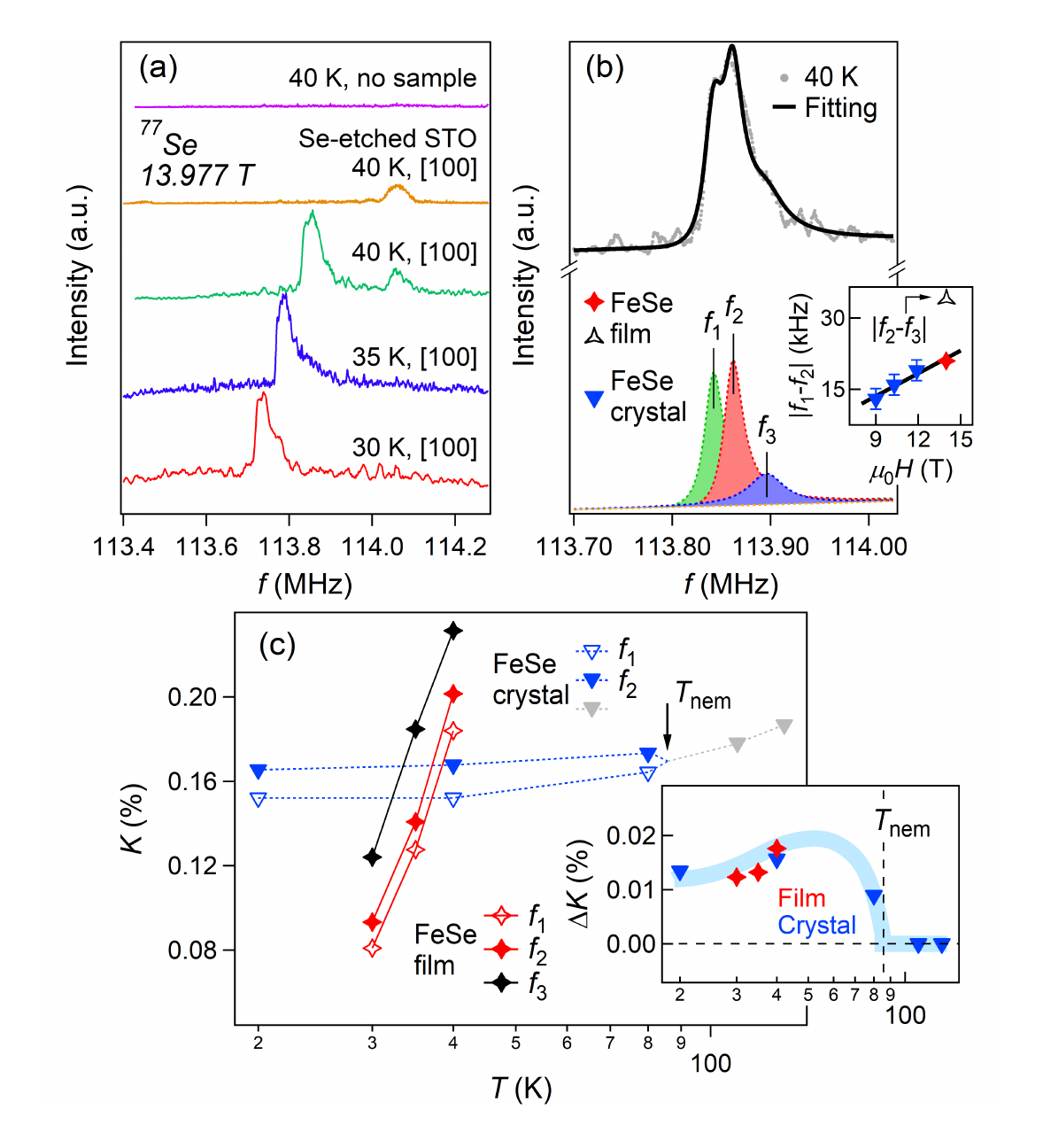}
  \end{center}
  \caption{\textbf{$^{77}$Se NMR measurements.}
  \textcolor{black}{(a) $^{77}$Se NMR spectra of 210-ML film and Se-etched SrTiO$_3$ substrate.}
  (b) \textcolor{black}{Multipeak fitting of the green curve in (a) by three Lorentzian peaks (dashed curves) with a linear background (dashed line).}
  Inset: Frequency splitting between $f_{\rm 1}$ and $f_{\rm 2}$ at 40 K as a function of field. Blue triangles represent the values of
  undoped FeSe crystals, adopted from Refs. [\cite{BaekSH2015,WangPS2017,LiJ2020PRX}].
  Black triangle indicates the splitting between $f_{\rm 2}$ and $f_{\rm 3}$ of FeSe film at 40 K.
  Thick line is a guide to the eye.
  (c) Temperature dependent $^{77}$Se NMR Knight shift of 210-ML film and FeSe$_{0.95}$S$_{0.05}$ crystal ($H$ $\parallel$ [100]).
  Inset: Temperature dependence of Knight shift splittings between $f_{\rm 1}$ and $f_{\rm 2}$ lines in the film and crystal.
  Translucent blue curve is a guide to the eye.
  }
\end{figure*}

To further clarify the domain structure of our films, we conduct NMR measurements on 210-ML film. Figure 2(a) shows the frequency-swept NMR spectra of the $^{77}$Se nuclei for a magnetic field of 13.977 T.
\textcolor{black}{At $T$ = 40 K, a prominent line ($\sim$113.85 MHz) and a much weaker signal ($\sim$114.06 MHz) are revealed on the sample for $H$ $\parallel$ [100], which are attributed to the FeSe film and Se-etched SrTiO$_3$ substrate, respectively [see Fig. S1 and Sec. 2 of Supporting Information for more  discussion on the assignment].}
Upon cooling, the spectra of FeSe film shift to lower frequencies with a decrease of the linewidth.
\textcolor{black}{But} it is difficult to observe a signal below 30 K and above 40 K, \textcolor{black}{this limited temperature range for a decent NMR signal is due to the longer spin-lattice relaxation time and smaller nuclear spin polarization at lower and higher temperatures, respectively (see more discussion in Sec. 3 of Supporting Information).}
By a closer look, \textcolor{black}{we notice that several components are contained in the spectra of FeSe film, including a double-peak splitting and a shoulder. The quantitative fit of the spectra in Figs. 2(b) and S2 (see details of the fitting procedure in Sec. 4 of Supporting Information) validate that each line consists of three Lorentzian peaks with central frequencies $f_{\rm 1}$, $f_{\rm 2}$, and $f_{\rm 3}$, unlike only two split peaks in the nematic phase of FeSe crystals (evidence for the twinned orthorhombic domains) \cite{BaekSH2015,WangPS2017,LiJ2020PRX}.}
In the inset of Fig. 2(b), we plot the \textcolor{black}{nematicity-related} frequency splitting of FeSe crystals at 40 K under different fields \textcolor{black}{(blue  triangle, adopted from Refs. [\cite{BaekSH2015,WangPS2017,LiJ2020PRX}])} together with $\mid$$f_{\rm 1}$ -- $f_{\rm 2}$$\mid$ \textcolor{black}{(red star)} and $\mid$$f_{\rm 2}$ -- $f_{\rm 3}$$\mid$ \textcolor{black}{(black triangle)} of \textcolor{black}{FeSe film} at 40 K. One
obtains that $\mid$$f_{\rm 1}$ -- $f_{\rm 2}$$\mid$ of FeSe film follows the linear field
dependence of frequency splitting of FeSe crystals, while $\mid$$f_{\rm 2}$ -- $f_{\rm 3}$$\mid$ noticeably deviates from \textcolor{black}{the trend.} \textcolor{black}{This implies that the $f_{\rm 1}$-$f_{\rm 2}$ splitting in FeSe film may arise from the twinned nematic domains like that in FeSe crystals. The $f_{\rm 3}$ line could then be attributed to a single-domain structure.}

In Fig. 2(c), we present the temperature evolution of $^{77}$Se Knight shift of FeSe film and FeSe$_{0.95}$S$_{0.05}$ crystal \cite{BaekSH2020FeSeS}. The Knight shift is determined from
$K$ = ($f_{\rm res}$ -- $\gamma$$B$)/$\gamma$$B$,
where $f_{\rm res}$ is the peak frequency of NMR spectrum, $\gamma$ = 8.13 MHz/T is the nuclear gyromagnetic ratio for $^{77}$Se, and $B$ is the external field. \textcolor{black}{Although the Knight shifts in the film show much stronger temperature dependence than that in the crystal (see more discussion in Sec. 5 of Supporting Information),} their Knight shift splittings between $f_{\rm 1}$ and $f_{\rm 2}$ lines at low temperatures,
$\Delta$$K$ = $K_{\rm 2}$ -- $K_{\rm 1}$ $\propto$ $\mid$$f_{\rm 1}$ -- $f_{\rm 2}$$\mid$,
are comparable and exhibit similar temperature behavior [inset of Fig. 2(c)].
\textcolor{black}{This similarity between film and crystal further demonstrates that the $f_{\rm 1}$-$f_{\rm 2}$ splitting in FeSe film is the signature of twinned domains associated with nematicity. Therefore, the single-domain ($f_{\rm 3}$) and twinned-domain ($f_{\rm 1}$-$f_{\rm 2}$) structures are revealed to coexist in 210-ML film.}
\textcolor{black}{The former one is reminiscent of the single-domained 15-ML film (see Sec. 6 of Supporting Information for the challenge of detecting its single-domain structure by NMR), where the SrTiO$_3$-induced tensile strain has been found to be sizeable and isotropic \cite{TanS2013}. We thus suggest that the single domain is most likely confined to a region near the interface with noticeable strain effect; accordingly, the twinned domains come from the bulk region further away from the interface with negligible strain effect. This assignment is further validated by the intensity contrast between $f_{\rm 3}$ and $f_{\rm 2}$ peaks [Figs. 2(b) and S2], whose intensity ratio (Table S1) can be regarded as the volume proportion between these two regions.}


\begin{figure*}[h!t]
  \begin{center}
  \includegraphics[width=10.5cm]{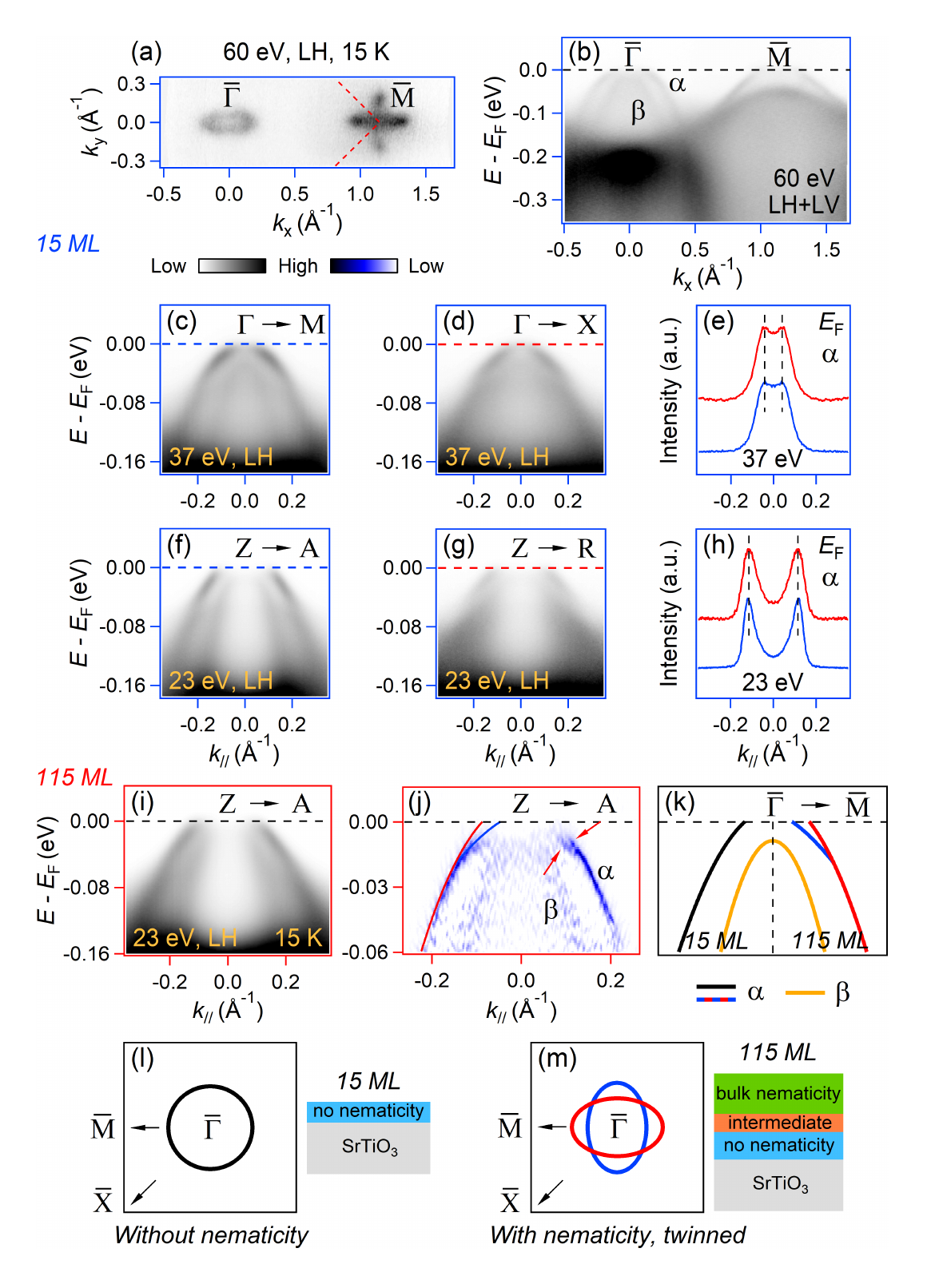}
  \end{center}
  \caption{\textbf{ARPES measurements of 15-ML and 115-ML films.}
  (a) Intensity plot at $E_F$ [$h\nu$ = 60 eV, linear horizontal (LH) polarization] of 15-ML film.
  (b) Intensity plot measured along the $\bar{\Gamma}$-$\bar{M}$ direction. The intensity is the sum of data acquired with linearly horizontal
  and vertical polarized photons.
  (c),(d) Intensity plots ($h\nu$ = 37 eV, LH polarization, $k_z$ = 0) recorded along the $\Gamma$-$M$ and $\Gamma$-$X$ directions, respectively.
  (e) MDCs of (c),(d) taken at $E_F$.
  (f)-(h) Same as (c)-(e) measured in $k_z$ = $\pi$ plane ($h\nu$ = 23 eV).
  \textcolor{black}{(i) Same as (f) of 115-ML film. (j) Curvature intensity plot of (i). Red arrows indicate the splitting of $\alpha$ band.
  (k) Low-temperature band structure cartoons of 15-ML and 115-ML films along the $\bar{\Gamma}$-$\bar{M}$ directions.}
  \textcolor{black}{(l)} Schematic low-temperature FS around $\bar{\Gamma}$ of single-domained FeSe without nematicity \textcolor{black}{(15-ML film).}
  \textcolor{black}{(m)} Same as \textcolor{black}{(l)} of twinned FeSe with nematicity \textcolor{black}{(115-ML film).}
  \textcolor{black}{Insets of (l) and (m) show the sketches of 15-ML and 115-ML films, respectively.}
  }
\end{figure*}

\textcolor{black}{Given that the domain structure in FeSe film evolves with the thickness, we study the electronic structures of 15-ML and 115-ML films by ARPES measurements.}
\textcolor{black}{It has been suggested that the vacuum-ultraviolet ARPES detects the signals of escaped photoelectrons mainly from the top two FeSe layers \cite{TanS2013}. This ensures the differences between 15-ML and 115-ML films observed below to be essentially of tensile-strain-effect origin.}
The FS mapping [Fig. 3(a)] and band dispersions along the $\bar{\Gamma}$-$\bar{M}$ direction [Fig. 3(b)] \textcolor{black}{of 15-ML film} are analogous to previous results of multilayer FeSe films \cite{TanS2013,ZhangY2016PRB}.
Due to the complicated band structure around $\bar{M}$ in FeSe, the orbital assignment of multiple 3$d$
bands and the energy scale of $d_{xz}$/$d_{yz}$ splitting at $\bar{M}$ are controversial \cite{NakayamaK2014,WatsonMD2015,ZhangP2015,ShimojimaT2014,YiM2019PRX,WatsonMD2016PRB,FedorovA2016,WatsonMD2017NJP,FanfarilloL2016,Borisenko2016NP,ColdeaA2018review}.
\textcolor{black}{Hereafter,} we focus on the relatively simple band structure around $\bar{\Gamma}$.
\textcolor{black}{Although the 15-ML film and detwinned FeSe crystal share the similarity of single-domain structure, there is no preferred orientation for the film  under SrTiO$_3$-induced tensile strain, distinct from the uniaxial strain mechanically added to the crystal \cite{YiM2019PRX,WatsonMD2017NJP,HuhSS2020,SuzukiY2015}. Thus, the former has tetragonal structure illustrated by STM while the latter shows orthorhombicity (see Sec. 7 of Supporting Information for discussion on the distinct NMR spectra caused by this structural difference).}
As a result, the ellipse-like hole pocket in Fig. 3(a) strongly contrasts with that in detwinned FeSe crystal below $T_{\rm nem}$ \cite{YiM2019PRX,WatsonMD2017NJP,HuhSS2020,SuzukiY2015} reflected in their completely different \textcolor{black}{photon polarization dependence} (see \textcolor{black}{Fig. S4} and \textcolor{black}{Sec. 8} of Supporting Information for more discussion).


We further measure the band structure along $\bar{\Gamma}$-$\bar{M}$ and $\bar{\Gamma}$-$\bar{X}$ directions in $k_z$ = 0 [Figs. 3(c)-3(e)] and $\pi$ [Figs. 3(f)-3(h)] planes \textcolor{black}{of 15-ML film.} As evidenced by the single set of peaks in momentum distribution curves (MDCs) along the $\bar{\Gamma}$-$\bar{M}$ directions [blue curves in Figs. 3(e) and 3(h)], no splitting is observed on \textcolor{black}{$\alpha$ band} when crossing $E_F$.
The hole pocket is further revealed to be circular in shape by comparing the $k_F$'s along $\bar{\Gamma}$-$\bar{M}$ and $\bar{\Gamma}$-$\bar{X}$ directions. In light of this, the ellipse-like FSs in Figs. 3(a) and \textcolor{black}{S4} are actually some segments of the circular pocket.
\textcolor{black}{In contrast, the FS of twinned FeSe crystals below $T_{\rm nem}$ consists of two crossed elliptical pockets at $\bar{\Gamma}$ arising from the split $\alpha$ band \cite{WatsonMD2015,FedorovA2016,Watson2015FeSeS}. The splitting is due to the dispersions of $d_{xz}$/$d_{yz}$ bands being different for the two twinned domains, and has therefore been suggested as a direct evidence for nematicity \cite{WatsonMD2015,WatsonMD2016PRB,FedorovA2016,Watson2015FeSeS,FernandesRM2014}.}
In a previous study of FeSe$_{1-x}$S$_x$ crystals \cite{Watson2015FeSeS}, as the reduction of $T_{\rm nem}$ by the increasing sulfur doping \cite{ColdeaAI2019}, the degree of FS anisotropy at $\bar{\Gamma}$ is lowered with reduced splitting of \textcolor{black}{$\alpha$ band.} It is expected that the
outer hole pocket would evolve into an isotropic circular FS when nematicity is quenched in FeSe$_{1-x}$S$_x$. Therefore, we suggest that
nematicity is absent in 15-ML film, \textcolor{black}{consistent with the tetragonal lattice symmetry therein.}

\textcolor{black}{As discussed above, the bulk region of thick FeSe films is composed of the nematicity-related twinned domains.
In Fig. 3(i), we measure the near-$E_F$ band structure of 115-ML film under the same conditions as in Fig. 3(f). To facilitate the visualization of
dispersive features, we present the two-dimensional curvature analysis \cite{PengsCurv} of the raw ARPES intensity in Fig. 3(j). As indicated by the red
arrows, the splitting of $\alpha$ band is clearly observed, its splitting energy scale of $\sim$20 meV (Fig. S5) is comparable with that in twinned FeSe crystals \cite{WatsonMD2015}. The presence of bulk-like nematicity in the bulk region of 115-ML film is compatible with the twinned domains therein. To illustrate the contrast between 15-ML and 115-ML films, we sketch their low-temperature band dispersions along the $\bar{\Gamma}$-$\bar{M}$ directions in Fig. 3(k), the corresponding FSs around $\bar{\Gamma}$ are shown in Figs. 3(l) and 3(m).}

\textcolor{black}{To further support the absence of nematicity in 15-ML film,} we study the splitting of $d_{xz}$ and $d_{yz}$ orbitals at $\bar{\Gamma}$, the anisotropy between which has been suggested as a hallmark of nematicity \cite{LeeCC2009,ChenCC2010}.
As shown in Figs. 4 and \textcolor{black}{S6,} we record the band structure \textcolor{black}{of 15-ML film} around $Z$ at different temperatures.
Upon warming, the \textcolor{black}{$\beta$ band} with band top at $\sim$--13 meV exhibits little change, as illustrated by the curvature intensity plots and energy distribution curves (EDCs) [Figs. 4(a), 4(b), \textcolor{black}{S6}]. The \textcolor{black}{$\alpha$ band} is also less temperature sensitive as evidenced by the nearly constant momentum values from the MDCs [Fig. 4(d)], which are taken at --40 meV to avoid the influence of thermal population effect. In Fig. 4(c), by parabolic fits to raw data at 30 and 150 K, we estimate the $d_{xz}$/$d_{yz}$ splitting as $\sim$22 meV for both temperatures, which is comparable with the spin-orbit-coupling-induced gap ($\sim$20 meV) determined in FeSe \cite{FedorovA2016,Borisenko2016NP}.
\textcolor{black}{The temperature-independent splitting dominated by spin-orbit coupling effect in 15-ML film is distinct from the 115-ML film, where the temperature evolutions of the $\alpha$ and $\beta$ bands are clearly revealed (Fig. S7).}



\begin{figure*}[h!t]
  \begin{center}
  \includegraphics[width=10.5cm]{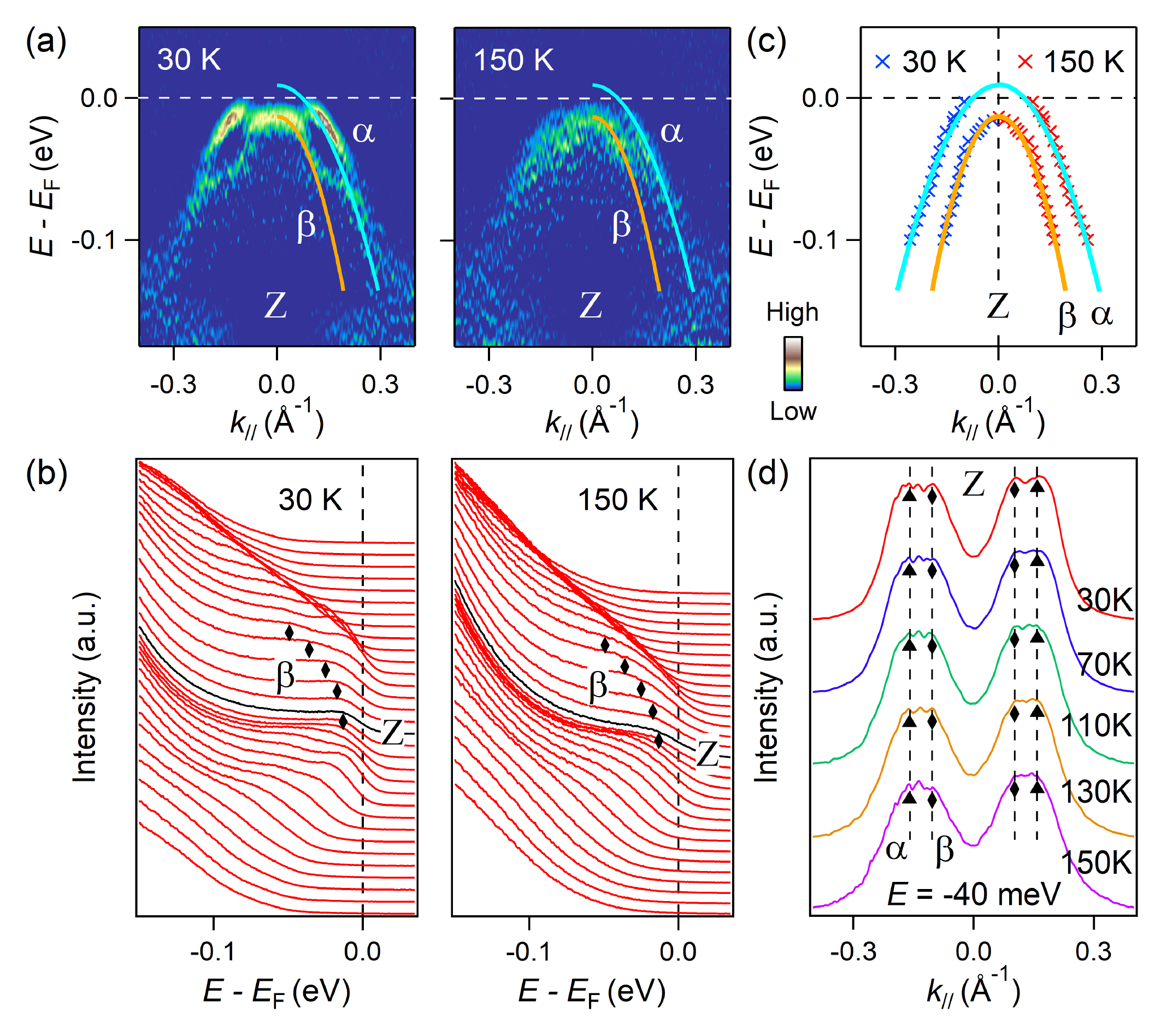}
  \end{center}
  \caption{\textbf{Temperature-independent $d_{xz}$/$d_{yz}$ splitting in 15-ML film.}
  (a) EDC curvature intensity plots ($h\nu$ = 23 eV, LH polarization) taken on 15-ML film along the $Z$-$A$ direction at 30 and 150 K.
  (b) Respective EDCs of raw data (--0.40 $\textless$ $k_{\rm \parallel}$ $\textless$ 0.40 $\AA^{-1}$). Diamond markers indicate the \textcolor{black}{$\beta$ bands.}
  (c) \textcolor{black}{Extracted dispersions of $\alpha$ and $\beta$ bands} at 30 and 150 K from EDCs and MDCs together. Solid curves are the parabolic fittings, which are also superimposed on (a).
  (d) Temperature-dependent MDCs taken at --40 meV. Triangle and diamond markers indicate \textcolor{black}{the $\alpha$ and $\beta$ bands,} respectively.
  }
\end{figure*}


\subsection{Discussion and conclusion}
\textcolor{black}{Our observations clearly demonstrate the tensile-strain-induced no nematicity near the interface, solving the aforementioned controversy about
the nematicity in multilayer FeSe/SrTiO$_3$ films \cite{TanS2013,Peng2015CPB,WenC2016NC,20ML2017PRB,XJZhou2016CPL}.}
The strain effect has been found to exhibit the exponential-like relaxation with increasing FeSe thickness \cite{TanS2013}.
\textcolor{black}{The nematicity strength is therefore expected to gradually recover towards the bulk value between the region without nematicity and the region with bulk-like nematicity, namely, an intermediate region probably exists. Overall, as illustrated in Fig. S8 [also the insets of Figs. 3(l) and 3(m)], we suggest that there are likely three regions.} In Sec. 12 of Supporting Information, we make an estimate on their spatial scales. Given that the volume of intermediate region is less than 10\% of the bulk region, the NMR lines with tiny intensities from the intermediate region could be hidden by the intense $f_{\rm 1}$-$f_{\rm 2}$ line.

\textcolor{black}{Last but not least, we discuss the implications of our results for the relationship between nematicity and superconductivity.
Extensive studies have been performed to uncover their interplay in FeSe crystals, but the findings are multifarious as described above, leaving it still rather mysterious.
\textcolor{black}{The absence of a superconducting gap for undoped multilayer (2--20 MLs) FeSe/SrTiO$_3$ films has been reported in previous STM studies \cite{WangQY2012,LiZ2014STM,ZhangT2016dose}.} Within this thickness range, the nematicity is not present either based on our results. \textcolor{black}{The simultaneous disappearance of nematicity and superconductivity in thinner FeSe films unambiguously suggests that the suppression of superconductivity therein
is not directly related to the nematicity.}}

\textcolor{black}{In summary, we have provided compelling evidence for the suppression of nematicity near the interface of multilayer FeSe/SrTiO$_3$ films. In the bulk region further away from the interface, where the SrTiO$_3$-induced tensile strain is fully relaxed, the bulk-like nematicity is revealed.} The present results facilitate further understanding the nematicity and its interplay with superconductivity in multilayer FeSe films.


\subsection{Methods}
The experimental methods and details of the film growth, ARPES measurements, STM measurements, and NMR measurements can be found in the Supporting Information.

\subsection{Data availability}
The data that support the findings of this study are available from the corresponding authors upon reasonable request.


\begin{acknowledgement}
We thank Anmin Zhang and Long Ma for stimulating discussions \textcolor{black}{and Danny Baumann for the fast and uncomplicated provision of the surface coils.}
This work was supported by the National Natural Science Foundation of China (Grant No. 11904144) and the Deutsche Forschungsgemeinschaft under Grant SFB 1143 (project C04). O.S., A.K., B.B. and S.B. acknowledge the support from the BMBF via project UKRATOP. R.L., B.B., S.B. and A.F. acknowledge
the support from the W{\"u}rzburg-Dresden Cluster of Excellence on Complexity and Topology in Quantum Matter--ct.qmat (EXC 2147, project-id
390858490).
\end{acknowledgement}

\subsection{Author contributions}
R.L. and O.S. contributed equally to this work.
R.L., S.B. and A.F. conceived the projects.
R.L., O.S. and A.F. synthesized the samples using MBE.
R.L. and A.F. performed ARPES measurements with the assistance of O.S. and A.K.
H.G. conducted NMR experiments.
M.K. performed STM measurements.
R.L., H.G., S.B. and A.F analysed the experimental data. R.L. wrote the manuscript with input from all the authors.

\subsection{Competing interests}
The authors declare no competing interests.

\subsection{Additional information}
\textbf{Supplementary information} is available in the online version of the paper.

\noindent \textbf{Correspondence} and requests for materials should be addressed to Rui Lou, Sergey Borisenko or Alexander Fedorov.

\end{document}